\documentclass[prb,12pt,onecolumn,superscriptaddress]{revtex4-1}

\usepackage{amssymb,graphicx,color,amsmath,xfrac,oldstyle,bm,setspace}
\usepackage[bookmarks,colorlinks=true,breaklinks,pdftitle={Resonant Ultrasound YBCO},pdfauthor={Arkady Shekhter}]{hyperref}  
\hypersetup{linkcolor=blue,citecolor=blue,filecolor=black,urlcolor=blue}

\newcommand{\ti}[1]{{#1}}
\newcommand{\YBCO}[1]{YBa$_2$Cu$_3$O$_{\mathrm{#1}}$}
\newcommand{\Tc}{T$_{\textrm{c}}$ }

\renewcommand{\Im}{{\mathrm{Im}}}
\renewcommand{\Re}{{\mathrm{Re}}}

\newcommand{\eps}{\varepsilon}

\newcommand{\xder}[2]{\left(\sfrac{\partial{#1}}{\partial{#2}}\right)}

\begin{document}

\title{Bounding the pseudogap with a line of phase transitions in \YBCO{6+\delta}.}

\author{Arkady~Shekhter}		\affiliation{Pulsed Field Facility, NHMFL, Los Alamos National Laboratory, Los Alamos, NM 87545}
\author{B.~J.~Ramshaw}			\affiliation{Pulsed Field Facility, NHMFL, Los Alamos National Laboratory, Los Alamos, NM 87545}
\author{Ruixing~Liang}			\affiliation{Department of Physics and Astronomy, University of British Columbia, Vancouver, BC, Canada, V6T 1Z1}
								\affiliation{Canadian Institute for Advanced Research, Toronto, Canada, M5G 1Z8}
\author{W.~N.~Hardy}			\affiliation{Department of Physics and Astronomy, University of British Columbia, Vancouver, BC, Canada, V6T 1Z1}
								\affiliation{Canadian Institute for Advanced Research, Toronto, Canada, M5G 1Z8}
\author{D.~A.~Bonn}				\affiliation{Department of Physics and Astronomy, University of British Columbia, Vancouver, BC, Canada, V6T 1Z1}
								\affiliation{Canadian Institute for Advanced Research, Toronto, Canada, M5G 1Z8}

\author{Fedor~F.~Balakirev}	\affiliation{Pulsed Field Facility, NHMFL, Los Alamos National Laboratory, Los Alamos, NM 87545} 
\author{Ross~D.~McDonald}	\affiliation{Pulsed Field Facility, NHMFL, Los Alamos National Laboratory, Los Alamos, NM 87545}
\author{Jon~B.~Betts}		\affiliation{Pulsed Field Facility, NHMFL, Los Alamos National Laboratory, Los Alamos, NM 87545}
\author{Scott~C.~Riggs}		\affiliation{Stanford Institute of Materials and Energy Sciences, Stanford University, Stanford, CA 94305, USA}
							\affiliation{Departments of Physics and Applied Physics, and Geballe Laboratory for Advanced Materials, Stanford University, Stanford, CA 94305, USA}
\author{Albert~Migliori}		\affiliation{Pulsed Field Facility, NHMFL, Los Alamos National Laboratory, Los Alamos, NM 87545}

%\begin{abstract}\end{abstract}
\date{\today}\maketitle
\setstretch{1.6} 
\setlength{\parindent}{0.cm}
\setlength{\parskip}{1cm}

\newpage

\textbf{ Close to optimal doping, the copper oxide superconductors show 'strange metal' behavior\cite{Transport,TransportHussey2009}, suggestive of strong fluctuations associated with a quantum critical point\cite{OrensteinMillis2000,vanderMarel2003,VarmaReports,MarginalFL1989}. Such a critical point requires a line of classical phase transitions terminating at zero temperature near optimal doping inside the superconducting 'dome'. The underdoped region of the temperature-doping phase diagram from which superconductivity emerges is referred to as the 'pseudogap'\cite{Timusk,NeutronsYBCO1,NeutronsYBCO2,NeutronsHg1201,Kaminski,VarmaPG,AjiVarma} because evidence exists for partial gapping of the conduction electrons, but so far there is no compelling thermodynamic evidence as to whether the pseudogap is a distinct phase or a continuous evolution of physical properties on cooling. Here we report that the pseudogap in \YBCO{6+\delta} is a distinct phase, bounded by a line of phase transitions. The doping dependence of this line is such that it terminates at zero temperature inside the superconducting dome. From this we conclude that quantum criticality drives the strange metallic behavior and therefore superconductivity in the cuprates.} 

Resonant ultrasound spectroscopy (RUS) measures the frequencies $f_n$ and widths $\Gamma_n$ of the vibrational normal modes of a crystal acting as a free mechanical resonator. The frequencies of the normal modes are determined by density and geometry of the crystal as well as its elastic properties.  The elastic component of the temperature evolution of these frequencies, $\Delta f_n(T)$, depends on a linear combination of all elastic moduli and reflects changes in the thermodynamic state of the system such as those associated with a phase transition. The width of a resonance $\Gamma_n(T)$ is proportional to the energy dissipation caused by time-dependent (dynamic) fluctuations in the system. Measuring many resonances provides access to elastic properties and fluctuations with different symmetries.\cite{RUS,Migliori-RSI,Migliori-YBCO,Birss} Recent advances in the quality of single crystal \YBCO{6+\delta} (YBCO) have pushed the boundary of possible measurements, as evidenced by the observation of quantum oscillations\cite{DoironLeyraud}. Advances in resonant ultrasound enable determination of the thermodynamics of these sub-millimeter crystals to part-per million accuracy. 

The narrow temperature range over which the resonances evolve across the superconducting transition illustrates the quality of the crystals and the accuracy of the measurement\cite{Bishop1987} (Figure~1). For the underdoped crystal, \YBCO{6.60}, we observe a sharp ($0.5$K wide) discontinuity in the resonance frequency, $\Delta{f}/f\approx10^{-4}$, at the superconducting transition (Figure~1). A sharper discontinuity is observed in the overdoped crystal, \YBCO{6.98}, a possible consequence of the reduction in oxygen disorder near optimal doping. The step discontinuity in resonance frequency and the accompanying discontinuous change (break) in slope are thermodynamic signatures of the superconducting transition (SI).

RUS measurements across the temperature range encompassing the pseudogap in the two YBCO crystals are shown in Figure~2. The temperature dependence of the resonance frequencies in underdoped \YBCO{6.60} reveals a break in slope at the pseudogap boundary $T^*=245K$---in itself a standard thermodynamic marker for a phase transition (Figure~2(a,c)). It differs from the signature of the superconducting transition in that there is no resolvable discontinuity in the frequency itself. This temperature is the same as the onset temperature of magnetic order observed by neutron scattering measurements of YBCO specimens of similar composition (Figure~3).\cite{NeutronsYBCO1,NeutronsYBCO2} In the overdoped crystal, \YBCO{6.98}, the break in slope of the temperature dependence is observed at $T^*=68$K, Figure~2(b). To emphasize the break in slope in these data, we use the redundant information contained in all observed resonances to extract the different contributions to their temperature dependences (Figure~4(c)). This process reduces the temperature dependence of all fifteen normal modes measured to three dominant components (see SI).  The blue and red curves in Figure~4(c) capture the effects of superconductivity and of fluctuations in the vicinity of the pseudogap, respectively.  The green curve, which has a break in slope at $T^*=68K$, corresponds to the thermodynamic effects at the pseudogap, revealing that the pseudogap occurs via a phase transition. 

The `strange metal' behavior that cuprates exhibit universally at higher temperature breaks down in the pseudogap region of the temperature-doping phase diagram,\cite{Timusk,TransportHussey2009,NMR,ARPES-AD,ARPES-AK,Nernst,LeridonMonod} where measurements indicate the presence of magnetic order\cite{Kaminski,NeutronsYBCO1,NeutronsYBCO2,NeutronsHg1201} (Figure~3). The break in slope that we observe in both underdoped and overdoped YBCO establishes the pseudogap as a thermodynamic phase that moves to lower temperature with increased doping. Observation of the pseudogap boundary below the superconducting transition temperature in overdoped YBCO indicates that the superconducting dome surrounds the zero-temperature end point of the pseudogap phase boundary. 

At both dopings the pseudogap is accompanied by a strong (up to hundred-fold in the overdoped crystal) increase in the width of the resonances at temperatures above the pseudogap phase boundary (Figure~2(c,d)). The width of the resonances are determined by the ultrasonic energy absorption (attenuation)\cite{Bhatia}, revealing strong fluctuations in the dynamics of the metallic state as it approaches $T^*$. From the width of the resonances we estimate the thermodynamic effects accompanying the pseudogap phase transition to be $\Gamma/f \sim 5\times10^{-3}$: about $50$ times larger than the relative modulus shift across the superconducting phase transition for both dopings. Energy absorption is highest when the measurement frequency matches the characteristic relaxation time of the system: $2\pi f \tau(T)=1$. $\tau$ diverges as the phase transition temperature is approached (critical slowing down)\cite{LandauKhalatnikov}, therefore the maximum in ultrasonic energy absorption is closer to the pseudogap phase boundary for resonances of lower frequency. For the underdoped crystal the width of the maximum and the contribution of the large phonon background at $245K$ obscures this effect. The overdoped crystal, with its narrower maxima and smoother background exhibits this effect clearly: $1/{\tau(T)}$ extrapolated from resonances at different frequencies vanishes at the pseudogap phase boundary (Figure~4(a,b)).  Causality requires that the maxima in energy absorption are accompanied by elastic stiffening over the same temperature range. This stiffening is observed in addition to the distinct break in slope at $T^*$ (Figure~2(b)). 

The potential for RUS to determine the broken symmetry in the pseudogap phase was limited in this study by the precision with which crystal shape could be controlled, an issue that may be resolvable as sample preparation techniques improve. The pseudogap phase transition is located by our RUS measurements with $\pm3$K uncertainty, improving on the $\pm30$K uncertainty in onset of neutron spin-flip scattering. This clearly separates the onset of  magnetic order\cite{Kaminski,NeutronsYBCO1,NeutronsYBCO2,NeutronsHg1201} at $T^*$ from the onset $T_K$ of the Kerr rotation signal\cite{Kerr} and charge order\cite{Xray} at lower temperature (Figure~3). In our measurements we observe an increase in energy absorption over a broad region near $T_K$ (Figure~2(c)), however we do not observe an accompanying thermodynamic signature there. Our observed evolution of the pseudogap phase boundary from underdoped to overdoped establishes the presence of a quantum critical point inside the superconducting dome, suggesting a quantum-critical origin for both the strange metallic behavior and the mechanism of superconducting pairing. 

%\begin{bibunit}[plain]
 
%\end{bibunit}

{\bf Acknowledgements} We thank  Elihu~Abrahams, James~Analytis, Philippe~Bourges, Alexander~Finkel'stein, Martin~Greven, Neil~Harrison, Kim~Modic, Chandra~Varma, Inna~Vishik, and Guichuan~Yu for critical reading of the manuscript and informative discussions. Work at Los Alamos National Laboratory (LANL) was supported by NSF-DMR-0654118, DOE, and the State of Florida. LANL is operated by LANS LLC. Work at the University of British Columbia was supported by the Canadian Institute for Advanced Research and the Natural Science and Engineering Research Council. This work was supported in part by the NSF under Grant No. PHYS-1066293 and the hospitality of Aspen Center for Physics. 

\newcommand{\newfig}[4]{ \begin{figure}[h!t]\includegraphics[width=#1\columnwidth]{#2}\caption{#4}  \label{#3} \end{figure} }

\newpage

\newfig{0.99}{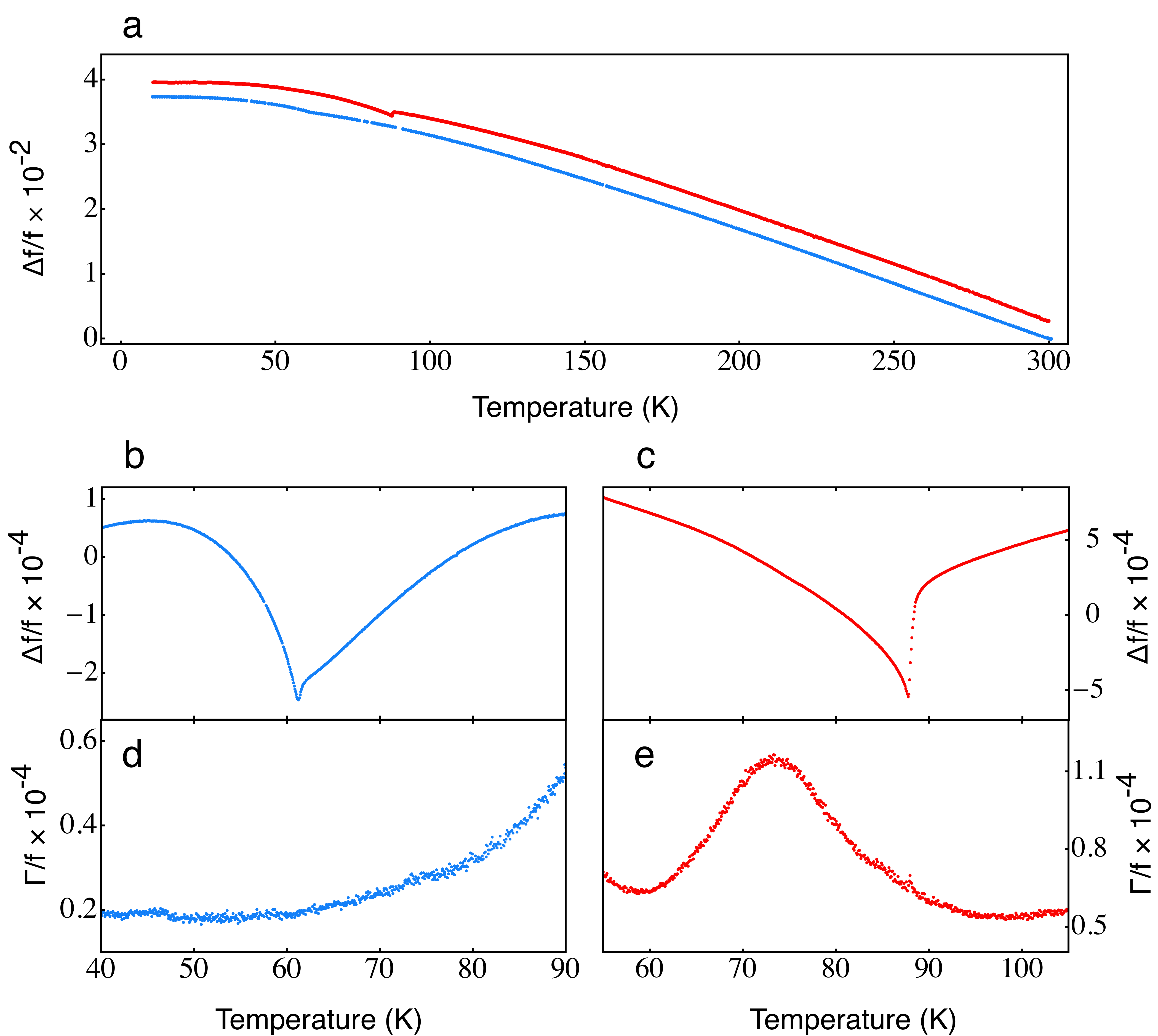}{fig:1}{
\textbf{Figure 1. The temperature evolution of resonances in underdoped and overdoped YBCO crystals. Superconductivity. } \textbf{a} A typical resonance frequency scan (normalized at room temperature) from room temperature to 10K for underdoped \YBCO{6.60} in blue with $T_c =61.6K$, and overdoped \YBCO{6.98} in red with $T_c =88K$. The scan for overdoped crystal is offset vertically for clarity. The smooth increase in frequency, which saturates at low temperature, is driven by the anharmonicity of the lattice and is typical of most solids.\cite{Varshni}. \textbf{b,c} Superconducting transition in the underdoped (b) and overdoped (c) crystals. Measurements were made at approximately $70mK$ steps. The elastic moduli drop discontinuously at the transition. The discontinuity is approximately one part in $10^{-4}$ in the underdoped crystal, and five parts in $10^{-4}$ in the overdoped. The form of the smooth monotonic background subtracted to obtain (b) and (c)was chosen only to emphasize the discontinuity\cite{Bishop1987}. \textbf{d,e} Resonance width for underdoped (d) and overdoped (e) YBCO. In the underdoped crystal no feature at the superconducting transition can be resolved. A broad maximum in resonance width well below $T_c$  in the overdoped crystal is an effect of the pseudogap (see text).}

\newpage

\newfig{0.99}{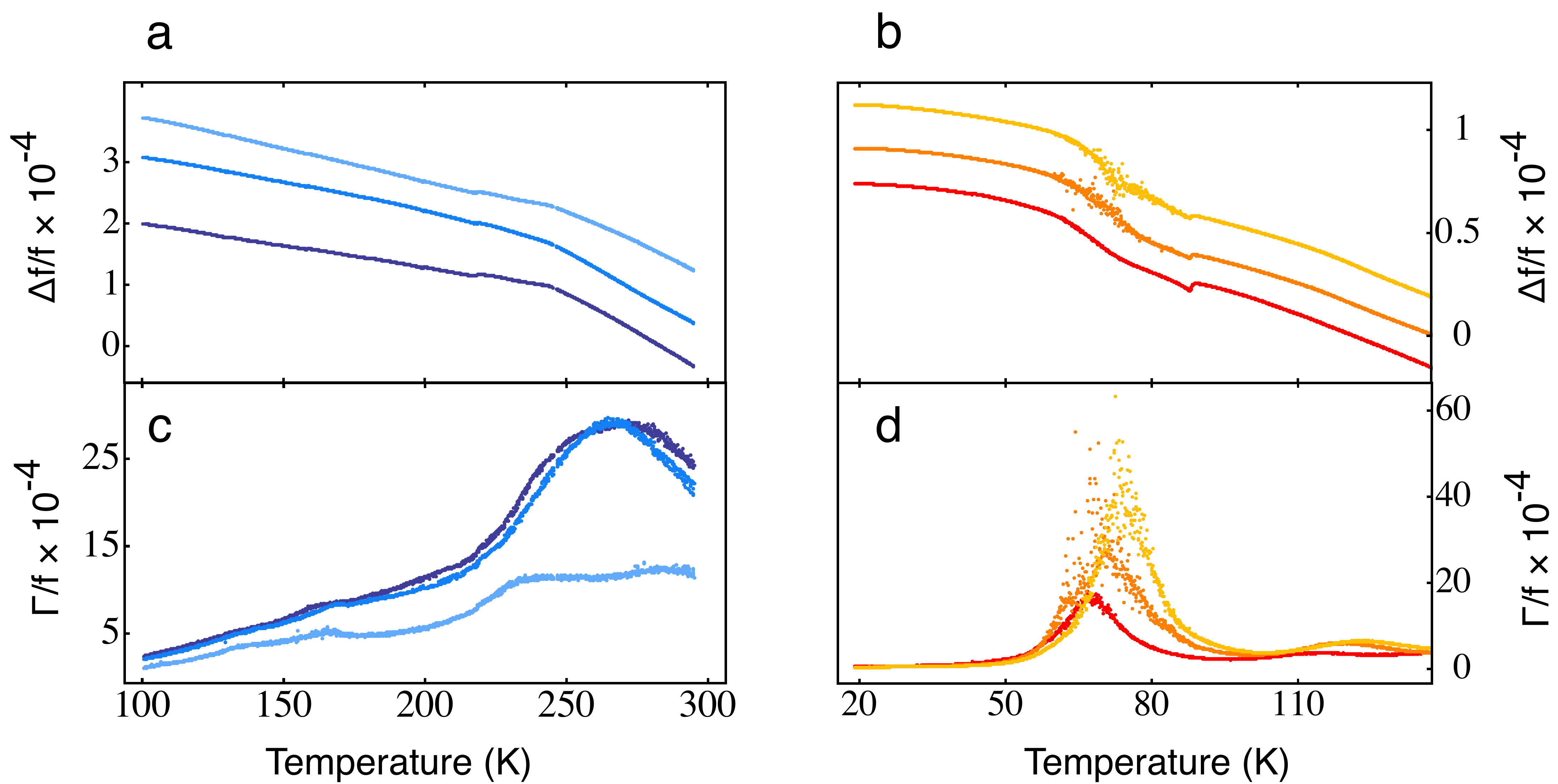}{fig:2}{
\textbf{Figure 2. The temperature evolution of resonances across the pseudogap phase boundary.} At both dopings a discontinuous change in slope of the temperature dependence of the frequency reveals a phase transition: underdoped (a) at $T^*=245K$, and overdoped (b) at  $T^*=68K$. At both dopings the resonance width has a broad maximum above $T^*$ (underdoped (c) and overdoped (d)). The break in slope is 5K wide in the underdoped crystal, 3K wide in overdoped. The increase in scatter of points near the break in slope in panel (b) is a result of strong increase in resonance width at this temperature, panel (d).}

\newpage

\newfig{0.8}{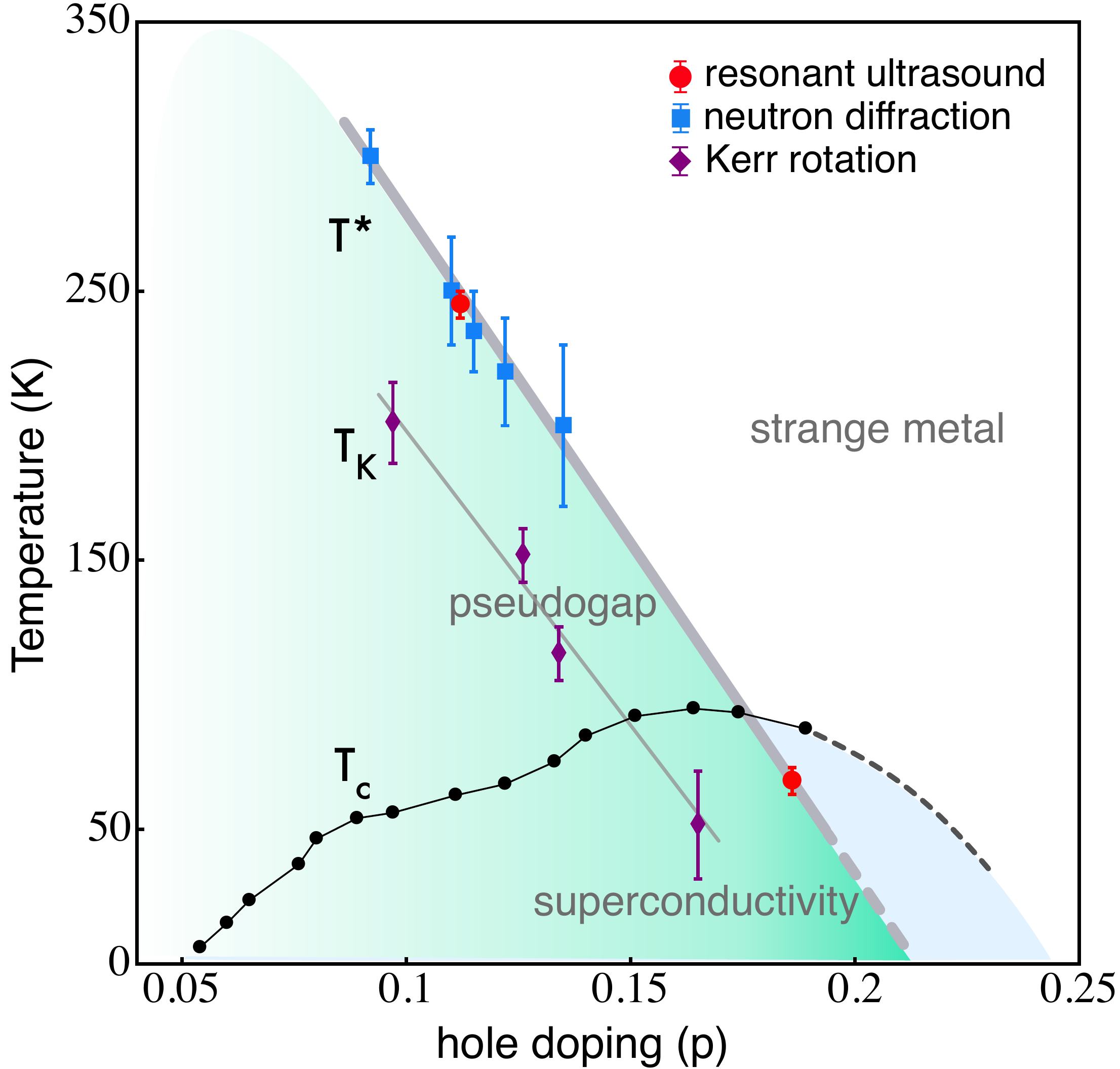}{fig:3}{
\textbf{Figure 3. The phase diagram of \YBCO{6+\delta}.} The pseudogap boundary in YBCO cuprates is indicated by a thick grey line (guide to the eye), as determined by neutron diffraction measurements\cite{NeutronsYBCO1,NeutronsYBCO2} (blue squares) and resonant ultrasound (red circles). The superconducting transition temperature is indicated by black circles\cite{LiangHardyBonn}. The temperature of the onset of Kerr rotation\cite{Kerr} where recent X-ray measurements detect an onset of charge order \cite{Xray} is shown in purple diamonds. Error bars represent the uncertainty in the determination of the onset temperature. The thin grey line is a guide for the eye. }

\newpage

\newfig{0.99}{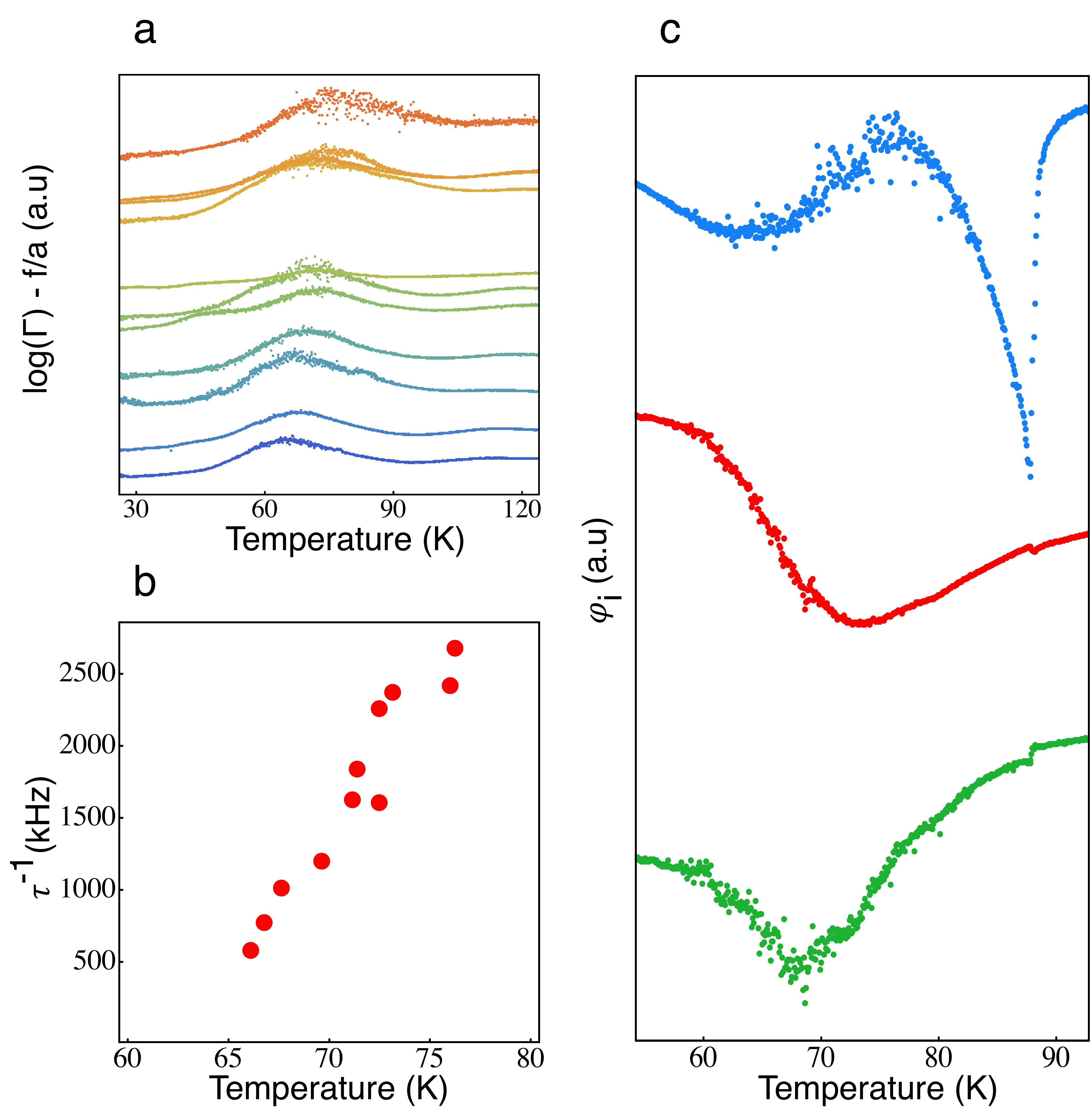}{fig:4}{
\textbf{Figure 4. The pseudogap boundary inside the superconducting dome.} \textbf{a} Evolution of resonance width with temperature across $T^*$ for several resonances.  To illustrate the evolution of the maximum in resonance width with resonance frequency each curve is offset vertically by an amount proportional to resonance frequency. \textbf{b} Evolution of the temperature of the resonance width maxima in \textbf{(a)} with resonance frequency ( $2\pi{f} \tau = 1$). The characteristic time $\tau$ increases as the pseudogap temperature is approached (critical slowing down)  \textbf{c.} Three different components of the temperature dependence of all resonance modes in the overdoped crystal: blue is dominated by superconductivity, red by fluctuations, and green by the pseudogap. The smooth anharmonic background, which dominates Figure 1(a), is not shown.  Each curve is scaled vertically for clarity.}

\newpage

\appendix

\centerline{\bf Supplementary information.} 
\vspace{1cm}

\section{Resonant ultrasound spectroscopy measurement system.}

The resonant ultrasound spectroscopy (RUS) measurement system comprises a piezoelectric driver and receiver, each in sufficiently weak point contact that the measured crystal acts as a free mechanical resonator. The essence of the measurement is the stress-stress response of the crystal: the driving transducer generates a stress at the point of contact on the crystal at a frequency $\omega$, and the receiving transducer generates a voltage proportional to the stress at different point on the crystal. The in-phase (real) and quadrature (imaginary) components of the voltage on the receiving transducer are recorded as a complex number $V(\omega)$. The measurement proceeds by sequentially changing the frequency of the driving transducer in a range that encompasses the lowest mechanical resonances of the crystal (typically about 20 resonances in the range 0.1 MHz to a few MHz for 1mm sized crystals).\cite{RUS-book} Generation of the driving signal and the phase-sensitive receiver logic are implemented on a custom-built electronic board  (similar to a heterodyne lock-in amplifier).\cite{Migliori-RSI} The shifts of the resonant frequencies with temperature, proportional to the shifts in the elastic moduli of the crystal, provide symmetry-specific information about the changes in the thermodynamic state of the system. The widths of the resonances are proportional to ultrasonic energy dissipation accompanying each resonance. 

High quality crystals are required so that the Q factor of the mechanical resonances are sufficiently high for this measurement. Detwinned crystals are necessary for this measurement because the motion of un-pinned twin boundaries in ultra-high purity crystals can be a large source of ultrasonic energy dissipation. This broadens the resonances to the point where they can not be resolved. Ultra-high-quality detwinned YBCO crystals are available in typical dimensions of 1x1x0.2mm, and sub-milligram mass. The small mass, small size, and plate geometry of the crystals require lower excitation power, better vibration isolation, and smaller transducer contact force than what is commercially available or previously used for RUS on larger crystals.\cite{alpha-Pu-reference}. Measuring a relatively large number of modes during a single temperature sweep that takes of order one week requires intelligent data acquisition; linear frequency sweeps across the whole range would take of order one year. These points define the experimental challenges. 

\begin{figure}[h!t]
\centerline{\includegraphics[width=0.6\columnwidth]{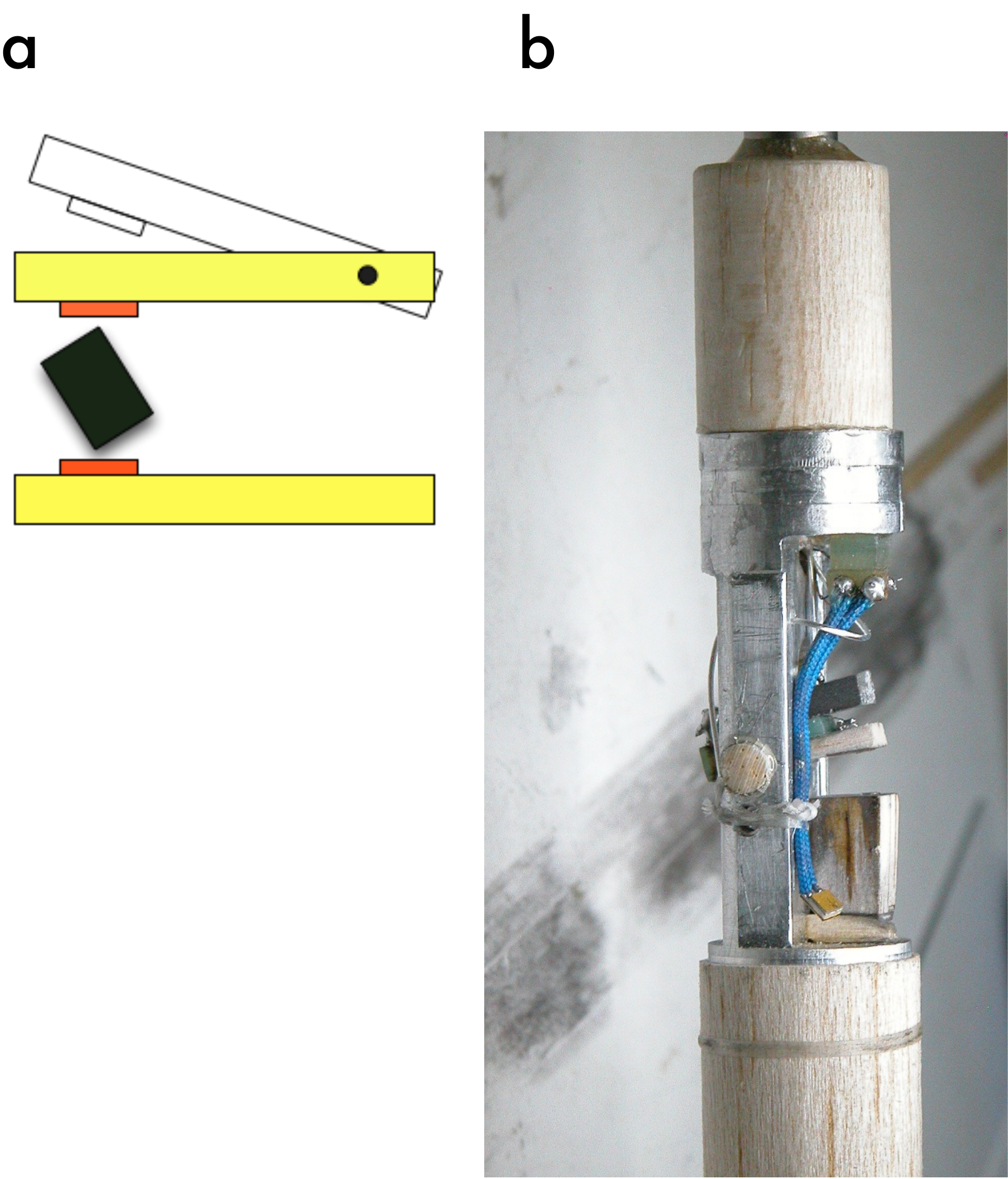}}
\caption{  (a) Geometry of the crystal-transducer assembly. The crystals are approximately $200\mu{m}$ thick, and one mm square.  Mounting the crystal on its corners ensures weak coupling to the crystal, allowing a free mechanical resonator conditions. (d) The RUS probe. Balsa-wood provides vibrational isolation over a broad temperate range, preventing acoustic crosstalk between the driving and pick-up  transducers. }
\label{fig:SI-1}
\end{figure}

Standard lithium niobate compressional mode transducers (disk-shaped, 1.5mm diameter, 200$\mu{m}$ thick) are used for both driver and receiver. To prevent vibrational crosstalk, the transducers must be acoustically isolated from one another and from the environment. Balsa-wood conveniently provides such isolation over a temperature range from liquid helium to room temperature. To obtain the sharpest resonances, the crystal was mounted on its corners, providing a single point of contact to each transducer and consistent coupling (Figure~\ref{fig:SI-1}). The ultrasonic power is adjusted as the temperature is swept to avoid non-linear effects and heating, while still maintaining the signal-to-noise ratio. By employing a non-uniform frequency scan, in which the frequency steps are small in the immediate vicinity of the resonance and large away from the resonance, we speed up (compared to a linear frequency sweep) the measurement across multiple resonances by about a factor of hundred while maintaining resonance frequency resolution exceeding one part per million. 

\subsection{Determining the resonance frequency and width using in-phase and quadrature components of the measured signal.}
\begin{figure}[h!t]
\centerline{\includegraphics[width=0.8\columnwidth]{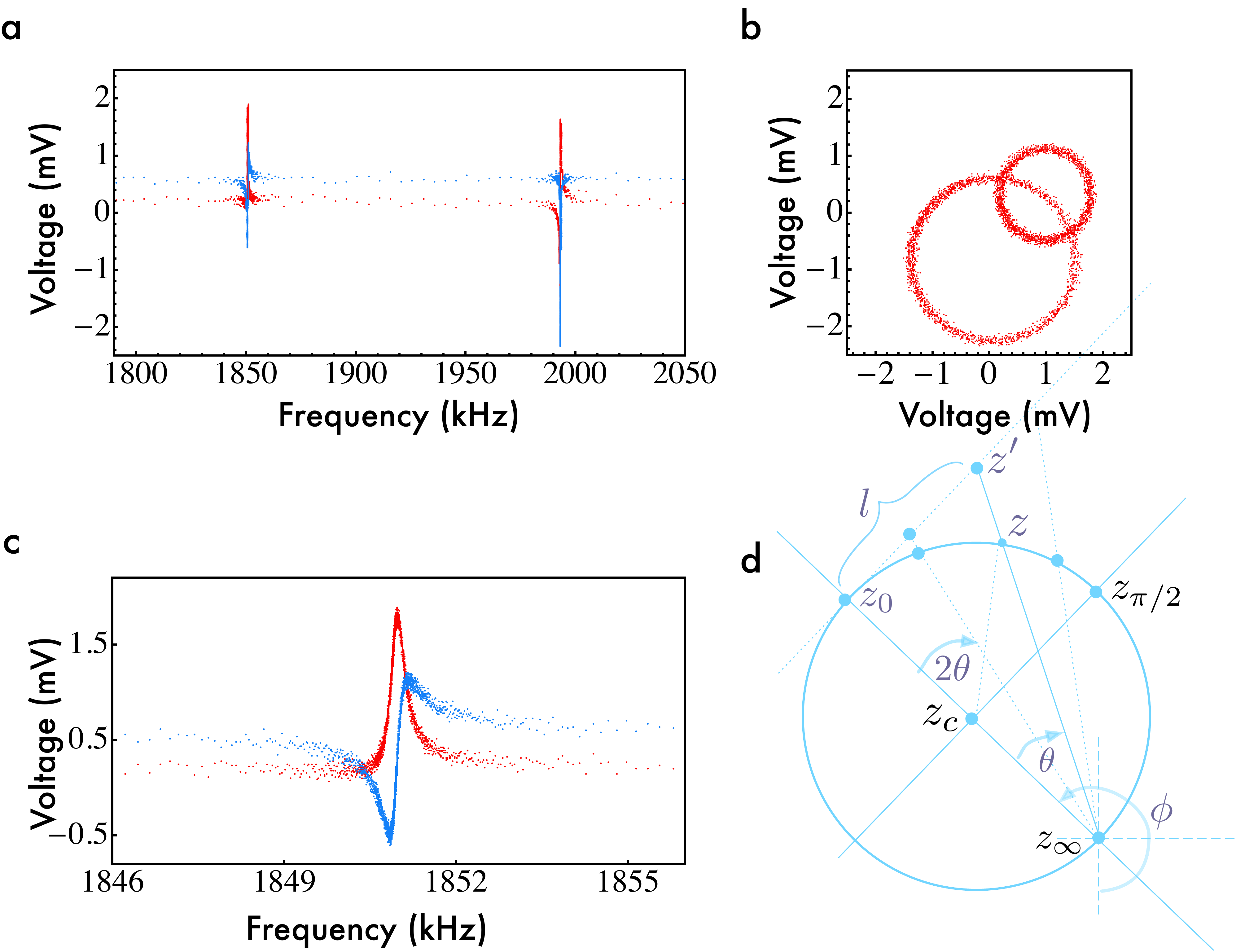}}
\caption{Frequency scan across the resonances and accurate determination of resonance frequency and width. (a)  A non-uniform scan in a broad frequency range encompassing two resonances: in-phase (red) and quadrature (blue) components of measured signal. (b) The same scan in the complex plane of voltage, i.e., quadrature and in-phase components on the receiving transducer are on the vertical and horizontal axes respectively. This panel illustrates a scan which is uniform in the complex plane of response voltage $V$: each data point is acquired at an equal time intervals (1ms), the frequency steps are adjusted in such a way that the complex response advances at a constant speed in the complex plane. (c) Scan across a single resonance illustrating the nonuniform-in-frequency scan in the close vicinity of the resonance. (d) All measured resonances have a Lorentzian shape, $z(\omega)= z_{\infty} + Ae^{i\phi}/(\omega-\omega_0 + i\Gamma/2)$. This panel illustrate basic geometrical facts that are necessary to accurately determine frequency and width of the resonance. A point $z(\omega)= z_{\infty} + Ae^{i\phi}/(\omega-\omega_0 + i\Gamma/2)$, in the complex $V$ plane traces a circle as we scan across a resonance, centered at $z_c$. The tails of the Lorentzian map to $z_{\infty}$, and the centre of the resonance at frequency $\omega_0$ maps to $z_0$.  To determine the resonance frequency and width, we use the identity $\theta = \arctan\big( (\omega-\omega_0)/(\Gamma/2) \big)$, where $\theta$ is the angle shown in the figure. A complementary strategy is to use the fact that the distance $\ell$ between the point $z'$, defined via $z'(\omega)-z_{\infty}=|z_0-z_{\infty}|^2/[z(\omega)-z_{\infty}]^*$, and the point $z_0$, is a linear function of frequency: $\ell=|z_0-z_{\infty}| (\omega-\omega_0)/(\Gamma/2)$. This reduces the determination of the resonance frequency and the width to a linear regression of all available data points. }
\label{fig:SI-2}
\end{figure}

The resonance measured by the pickup transducer has a Lorentzian shape, $V(\omega) = z_{\infty} + Ae^{i\phi}/(\omega-\omega_0 + i\Gamma/2)$, where $\Gamma$ is the resonance width and $z_{\infty}$ is the background in the vicinity of the resonance.  The resonance frequency and the width can be determined by  finding the frequency of the maximum amplitude of the signal and the width at half maximum. However, in measurements with marginal signal-to-noise, or in this case very weak coupling to maintain free mechanical vibration, this procedure does not produce reliable frequencies and widths.  Our approach is to use both the amplitude and the phase information of the signal $V(\omega)$, and then fit the data to a circle in the complex plane (Figure~\ref{fig:SI-2}). This first requires finding the centre of the circle, $z_c$, and then the resonant frequency and the width are fit via regression of all available data points in the vicinity of the resonance (as described in the caption of Figure~\ref{fig:SI-2}.) The center of the circle is found by using a variant of a Hough transform: circular waves emanating from each data point in the complex plane will interfere constructively at the center of the circle, and destructively elsewhere. This procedure is robust against noise and distortion of the circle caused by transients and non-linear effects.

\subsection{Understanding the effects of the superconducting and pseudogap physics on the observed frequency shifts.}
The temperature dependence of the resonance frequencies is determined by a superposition of the effects of several physical processes, each with a distinct temperature dependence of its own, for example superconductivity and the pseudogap. The problem of extracting the dependence of these two components from about 15 measured frequencies is that of solving 15 linear equations for only 2 or 3 unknown temperature dependent contributions. This overdetermination makes it possible to extract the evolution of the effects of pseudogap and superconductivity, as shown in Figure~4(d) in the main text. The problem is to find the $D$ functions $\phi_{i}(T)$ that capture the temperature dependence of $N>D$ frequencies with minimal error, $\Delta{f_n(T)}= \sum\limits_{i=1..D} \beta^n_i\phi_{i}(T)$. Here the temperature $T$ is a vector index. This defines the functions  $\phi_i(T)$  as the first $D$ eigenvectors of the matrix $M_{T,T'} = \sum\limits_{n=1..N} \Delta{f_n(T)} \Delta{f_n(T')}$ with the largest eigenvalues. The expansion coefficients $\beta_i^n$ can subsequently found by expanding vectors $\Delta{f_n(T)}$ in the basis of ${\phi_i(T)}$. The benefit of this approach is that you do not need to identify the mode corresponding to each resonance frequency, however, as a result this does not reveal the symmetry of the separate physical processes. 

\section{Background on the thermodynamic and transport phenomena associated with the elastic response.}
To examine the ultrasonic signature across a phase transition (discontinuities in the resonance frequency and the attenuation response), we present the elastic strain and stress in terms of the irreducible representations of the underlying lattice. For simplicity, we illustrate this point for a tetragonal system, the weak orthorhombicity of YBa$_2$Cu$_3$O$_{6+\delta}$ can then be introduced as a small perturbation. 

The six components of the strain tensor in the tetragonal lattice fall into five irreducible representations of the lattice point group: two compressional strain components, $\eps_{xx}+\eps_{yy}$ and $\eps_{zz}$, are crystallographic scalars belonging to the $A_{1g}$ representation.\cite{Birss}; four shear strain components,  $\eps_{xx}-\eps_{yy}$, $\epsilon_{xy}$, and a pair $(\eps_{xz},\eps_{yz})$,  fall into three non-trivial irreducible representations,  $B_{1g}$,$B_{2g}$ and $E_g$, respectively. We define the index $m=A_{1g},B_{1g}, B_{2g}, {E_g}$, and use it to label the components of strain, stress, and the elastic moduli.  As an example, in the presence of a (scalar) order parameter $\eta$, the free energy can be written as $d\mathbb{F}(T,\eta,\eps_m) = -SdT + \phi d\eta + \sigma_m d\eps_m $, where $\phi$ is the thermodynamic conjugate to the order parameter $\eta$---a restoring force equal to zero in thermodynamic equilibrium---and $\sigma_m$ is an elastic stress.  

Elastic deformations perturb the local thermodynamic equilibrium of the crystal, resulting in a coupling to thermodynamic variables such as temperature or order parameters. Crystallographic scalar components of strain can couple linearly to temperature and to order parameters. In this notation the elastic moduli are $\lambda_{mn} = \xder{ \sigma_m}{\eps_n}_{T}$.  In a tetragonal crystal, the linear couplings of elastic strain to temperature $\delta{T}$, and to a scalar order parameter $\delta\eta$, are controlled by the thermodynamic coefficients $\beta_m = -\xder{ \sigma_m}{T}_{\eps_m}$ and $Z_m = \xder{\sigma_m}{\eta}_{T,\eps}$ via $\Delta \mathbb{F}= -\delta{T}\beta_m\epsilon_{m} + \delta\eta Z_m  \epsilon_{m}$, where $Z_m$ and $\beta_m$ are only non-zero in the scalar crystallographic symmetry channels. We note that while shear strains are non-scalar in a tetragonal crystal, they can be scalar in a lower symmetry environment. For example, the weak orthorhombicity of the YBCO crystal structure (characterized by a small $B_{1g}$ distortion) introduces a small linear coupling of $B_{1g}$ shear strain to heat and scalar order parameters via a scalar  $\zeta_o (\epsilon_{xx} - \epsilon_{yy} ) $, where  $\zeta_o$ is the magnitude of distortion. In addition, for non-scalar order parameters, the shear strain can couple linearly in some situations. For example, a polar magnetic vector  \cite{AjiVarma,ShekhterVarma}, $\bm{\eta} = (\eta_x,\eta_y)$ can couple linearly to shear elastic strain in the $B_{1g}$ and $B_{2g}$ symmetry channels via order parameter bilinears $B_{1g} = \eta_x^2-\eta_y^2$ and $B_{2g} = \eta_x\eta_y$.

When elastic strain is coupled to the dynamics of the order parameter or to heat flow, the elastic response acquires a frequency (and momentum) dispersion. This can be described by introducing dynamic elastic response functions: $\lambda_{mn}(\omega) = \lambda_{mn}+R_{mn} A(\omega)$ where $R_{mn}$ is the difference between the fast and slow values of the elastic moduli, and $A(\omega)$ is proportional to the full dynamic correlation function of the physical process in question, such as heat flow or order parameter fluctuation.   $A(\omega)$ is normalized such that $A(\omega\rightarrow\infty)=1, \quad A(\omega\rightarrow0)=0$, and must be analytic in the upper half of the complex plane of $\omega$.  For example, in the case of coupling elastic strain to heat, $R_{mn}$ is equal to the difference between the adiabatic (fast) and isothermal (slow) elastic moduli:\cite{Bhatia} $R_{mn}=\left(\lambda_{mn}\right)_{S}-\left(\lambda_{mn}\right)_{T}=T\beta_m\beta_n/C_{\eps}$, where $C_{\eps}$ is the heat capacity at constant strain, and $\beta_m$ is the inverse of the thermal expansion coefficient (defined in the previous section). When elastic strain couples to the dynamics of an order parameter, $R_{mn} = \left(\lambda_{mn}\right)_{\eta} - \left(\lambda_{mn}\right)_{\phi} = {Z_mZ_n}/{Y}$ where $Y=\xder{\phi}{\eta}$ is the order parameter stiffness, and $Z_m$ is the coupling coefficient between order parameter and stress as defined above. In a tetragonal crystal, $R_{mn}$ is only non-zero for compressional strains. 

In a tetragonal crystal the change in the compressional elastic moduli across a second order phase transition is equal to $\left(\lambda_{mn}\right)_{\eta} - \left(\lambda_{mn}\right)_{\phi} = {Z_mZ_n}/{Y}$. Near the phase transition $Z\propto T_c\eta_0$, and $Y\propto T_F\eta_0^2$, which gives an estimate for the jump in elastic moduli (on warming) across the transition, $\delta\lambda \propto T_c^2/T_F$. Thus the relative change in frequency $\delta{f}/f_0$, which is proportional to the relative change in elastic moduli, is estimated as $\delta\lambda/\lambda\propto (T_c/T_F)^2$, where we assume that compressional elastic moduli in a metal are within a factor of order unity equal to  the Fermi energy per unit cell volume. This estimate compares well with the observed jump across the superconducting transition in YBCO, equal to $\Delta{f}/f\sim(T_c/T_F)^2\sim10^{-4}$, where $T_F$ is the Fermi temperature which is 5000K. 

Near a phase transition the order parameter can have relaxational dynamics\cite{LandauKhalatnikov,Kinetics}; the rate of relaxation of a perturbed order parameter is proportional to the order parameters restoring force $\phi$: ${d\delta\eta}/{dt}= -\gamma \phi = -\gamma (\delta F/\delta\eta)$. In this conventional case, the function $A(\omega)$ has the form $A(\omega) = \sfrac{-i\omega}{-i\omega + \frac1{\tau}}$  where $\tau=1/(\gamma Y)$. The frequency dispersion of the elastic response hence shifts the resonant frequencies and gives them a width proportional to real and imaginary parts of $A(\omega)$, 
\begin{align}\label{eq:KK}
\frac{\Gamma}{\omega_0} 
=& -R_s \Im A(\omega_0) \notag\\
\frac{\delta\omega}{\omega_0}  
=& \quad R_s \Re A(\omega_0) \, .
\end{align}
Here $R_s$ is of order of $R_{mn}/\lambda_{mn}$ and depends on the geometry of a normal mode associated with the resonance.\cite{Bhatia} For example, for magnetic phase transitions in solids $R_s$ is few percent.\cite{Bhatia} At $T^*$ the specific form of $A(w)$ is not as simple as in our example, but Eq.~\ref{eq:KK} still applies.

The width $\Gamma$ in Eq.~\ref{eq:KK} is the width of a resonance measured in the RUS experiment, and is proportional to the energy dissipation that accompanies the vibrational mode associated with that resonance.\cite{LandauVol5} Attenuation typically reaches a maximum when the timescale of the fluctuations match the frequency of the measured resonance. As the system approaches a phase transition, the relaxation time $\tau$ becomes very long (critical slowing down) and the condition $\omega \tau = 1$ can be satisfied for ultrasonic frequencies.

%\begin{bibunit}[plain]
 
%\end{bibunit}

\newpage

\end{document}